\documentclass[aps,prl]{revtex4}
\usepackage{epsfig}

\newcommand{\be}{\begin{equation}}
\newcommand{\ee}{\end{equation}}

\newcommand{\bmath}{\begin{mathletters}}
\newcommand{\emath}{\end{mathletters}}

\begin{document}

\title{\Large{\bf COEXISTENCE OF FERROMAGNETISM WITH SPIN TRIPLET SUPERCONDUCTIVITY}}

\vskip0.5cm 

\author{ {\bf Proceedings of the European Conference ''Physics of Magnetism '08''}\\ \vskip0.5cm
Grzegorz G\'{o}rski, Krzysztof Kucab\footnote{Corresponding author. E-mail: kkucab@univ.rzeszow.pl (K.\ Kucab)} and Jerzy Mizia }

\address{Institute of Physics, Rzesz\'{o}w University, Al.\ Rejtana 16A, 35-959 Rzesz\'{o}w, Poland\\}

\vskip0.5cm 


\begin{abstract}

\noindent

The experimental results for ZrZn$_{2}$, URhGe, and in some pressure ranges also for UGe$_{2}$,  have shown that the ferromagnetic superconductors are weak itinerant ferromagnets. Guided by these results we describe the phenomenon of coexistence between equal spin triplet pairing superconductivity (SC) and ferromagnetism (F) using the extended Stoner model, which includes in Hamiltonian the on-site Coulomb interaction, $U$, and occupation dependent hopping integral. We use Hartree-Fock (H-F) approximation and the Green functions technique. In the H-F approximation the on-site Coulomb interaction plays the role of the on-site exchange (Hund's) field. All inter-site interactions will have included the inter-site kinetic correlation, $\langle c_{i\sigma}^{+}c_{j\sigma}\rangle $, within the H-F approximation. We introduce the pressure-dependence to the hopping integral.  Numerical results are compared with experimental data for ZrZn$_{2}$.\ The kinetic correlation creates the superconductivity without help of negative values of Coulomb interactions. The model can explain stimulation of triplet SC by the weak itinerant ferromagnetism. This effect was observed experimentally in ZrZn$_{2}$. Numerical analysis also confirms the experimental effect of decrease in critical temperatures (Curie and superconducting) with increasing external pressure. %
\end{abstract}
\maketitle

%
\section{1. Introduction}    
The theoretical possibility of ferromagnetism (F) coexisting with the triplet parallel spins superconductivity (SC) was suggested by Fay and Appel for ZrZn$_{2}$ \cite{Fay1}. Further theoretical development took place after finding experimental evidence for coexistence of triplet SC with F (see e.g.\ \cite{Powell1}). 
It is relatively recently that the so-called ferromagnetic superconductors have been discovered, which at some high pressures exhibit the ferromagnetic and a spin triplet superconducting phases at the same time. At present UGe$_{2}$ \cite{Saxena1}, URhGe \cite{Aoki1} and ZrZn$_{2}$ \cite{Pfleiderer2} belong to the ferromagnetic superconductors. In ZrZn$_2$ compound the ambient pressure strength affects the Curie temperature. This compound has the quasi-linear dependence of both magnetic moment and Curie temperature on pressure.

\section{2. The Model Hamiltonian}

Taking into account that the ferromagnetic superconductors are weak itinerant ferromagnets we can describe them by the extended Stoner model (see e.g.\ \cite{Mizia1}), which includes the on-site Coulomb repulsion, $U$, and occupation dependent hopping integral. We can write the model Hamiltonian as
\begin{eqnarray}
H =  - \sum\limits_{ < ij > \sigma } {\left[ {t_0 - \Delta t\left( {\hat n_{i - \sigma }  + \hat n_{j - \sigma } } \right) + 2t_{ex} \hat n_{i - \sigma } \hat n_{j - \sigma } } \right]c_{i\sigma }^ +  c_{j\sigma } }  \;\;\;\;\;\;\;\;\;\;\;\;\;\;
\nonumber \\
 -\mu \sum\limits_i {\hat n_i }  + \frac{U}{2}\sum\limits_{i\sigma } {\hat n_{i\sigma } \hat n_{i - \sigma } }, 
\label{eq_1}
\end{eqnarray}
where $\mu$ is the chemical potential. The $\Delta t$ and $t_{ex}$ terms correspond to hopping and exchange-hopping interaction respectively, and are given by
\begin{equation}
\Delta t=t_0 - t_1 = t_0 \left( {1 - S_1 } \right),\;\; t_{ex} = \frac{t_0 + t_2}{2}-t_1 = \frac{1}{2}t_0 \left( {1 + S_1 S_2  - 2S_1 } \right),
\label{eq_2}
\end{equation}
where $S_1  \equiv {{t_1 } \mathord{\left/ {\vphantom {{t_1 } {t_0 }}} \right. \kern-\nulldelimiterspace} {t_0 }}$, and $S_2  \equiv {{t_2 } \mathord{\left/ {\vphantom {{t_2 } {t_1 }}} \right.
 \kern-\nulldelimiterspace} {t_1 }}$.
In equations above, $t_0$, $t_1$, and $t_2$ are the hopping amplitudes for an electron with spin $\sigma$ when both sites $i$ and $j$ are empty, when one of the sites $i$ or $j$ is occupied by an electron with opposite spin, and when both sites $i$ and $j$ are occupied by electrons with opposite spin, respectively.

In the calculations below, the constants $t_0$, $S_1$ and $S_2$ will be assumed pressure-dependent. The kinetic interactions $\Delta t$ and $t_{ex}$ will also depend on pressure through Eq.\ (\ref{eq_2}). The on-site Coulomb repulsion, $U$, will be assumed pressure-independent.
Taking into account the results of \cite{Lo1} giving us the dependence of effective mass, $m^*$, on the pressure, $p$, and comparing the dispersion relation in the tight binding approximation (at small $k$) with the expression $\varepsilon _k  = {{\hbar ^2 k^2 } \mathord{\left/
 {\vphantom {{\hbar ^2 k^2 } {2m^* }}} \right.
 \kern-\nulldelimiterspace} {2m^* }}$, one obtains the relation
\begin{equation}
t\left( p \right) = \frac{{t_0 }}{{1 - Ap}},
\label{eq_3}
\end{equation}
where for ZrZn$_2$ $A = 0.017\pm0.004\;{\rm kbar}^{{\rm  - 1}}$, and the lattice constant $a = 7.393$\AA $\;$(see \cite{Pfleiderer2}). Further on the dependence of hopping integrals on pressure will be suppressed in the notation, i.e.\ $t(p)\equiv t$, $\Delta t(p)\equiv \Delta t$, $t_{ex}(p)\equiv t_{ex}$.

In Hamiltonian (\ref{eq_1}) there are terms with four and six operators. The terms with four operators will be approximated by the average of two of them multiplied by the remaining two. The averages of the spin-flip type, $\left\langle {c_{i\sigma }^ +  c_{j - \sigma } } \right\rangle$, will be ignored. The six-operator term standing at $t_{ex}$ will be approximated by the product of two averages of two operators multiplied by the remaining two operators (see \cite{Arrachea1}). Using these approximations and dropping in the Hamiltonian spin singlet term and opposite spin triplet term we obtain, after transforming into momentum space, the following form
\begin{equation}
H = \sum\limits_{k\sigma } {\left( {\varepsilon _k^\sigma   - \mu + M^\sigma  } \right)\hat n_{k\sigma } }  - \sum\limits_{k\sigma } {\left( {\Delta _k^\sigma  c_{k\sigma }^ +  c_{ - k\sigma }^ +   + \rm{h.c.}} \right)},
\label{eq_4} 
\end{equation}
where $\varepsilon _k^\sigma   = \varepsilon _k b^\sigma$ is the spin-dependent modified dispersion relation, with the bandwidth factor, $b^{\sigma}$, given by
\begin{eqnarray}
b^\sigma   = 1 - \frac{2}{t}\left[ {\Delta tn_{ - \sigma }  - t_{ex} \left( {n_{ - \sigma }^2  - I_{ - \sigma }^2  - 2I_\sigma  I_{ - \sigma } } \right)} \right].
\label{eq_5}
\end{eqnarray}
The spin-dependent modified molecular field, $M^\sigma$, is expressed as 
\begin{equation}
M^\sigma   = Un_{-\sigma} + 2zI_{ - \sigma } \left( {\Delta t - 2t_{ex} n_\sigma  } \right), \;\;\;\;\;\;\; I_\sigma=\langle c_{i\sigma}^{+}c_{j\sigma}\rangle,
\label{eq_6}
\end{equation}
where $z$ is the number of nearest-neighbors.
The equal spin pairing (ESP) parameter, $\Delta_k^\sigma$, for the two-dimensional square lattice is given by
\begin{equation}
\Delta _k^\sigma   = 4t_{ex} I_{ - \sigma } \left( {\Delta ^\sigma _x \sin k_x  + \Delta ^\sigma _y \sin k_y } \right), \;\;\;\;\;\Delta ^\sigma _{x\left( y \right)}  = \left\langle {c_{i + x\left( y \right)\sigma } c_{i\sigma } } \right\rangle.
\label{eq_7}
\end{equation}

Solving the Green's function equations of motion (see \cite{Mizia1}) with Hamiltonian (\ref{eq_4}) we obtain the following relation for the ESP superconducting critical temperature
\begin{equation}
1 = 4t_{ex}I_{-\sigma}\frac{1}{N}\sum\limits_k {\left (\frac{{ {\sin ^2 k_x}  }}{{E_k^\sigma  }}\tanh \frac{{E_k^\sigma  }}{{2k_B T}}\right )}, 
\label{eq_8}
\end{equation}
where $E_k^\sigma   = \sqrt {\left( {\varepsilon _k^\sigma   - \mu + M^\sigma  } \right)^2  + \left( {2\Delta _k^\sigma  } \right)^2 }$, $T$ is the temperature, and $k_B$ is the Boltzmann's constant.

\noindent The equations for carrier concentration, $n$, and magnetization, $m$  

\begin{equation}
n = n^\sigma   + n^{ - \sigma }, \;\;\;\;\;\;\;\;\;\; m = n^\sigma   - n^{ - \sigma }
\label{eq_8a}
\end{equation}
can be obtained from $n^\sigma$

\begin{equation}
n^{ \sigma }  = \frac{1}{2}\left[ {1 - \frac{1}{N}\sum\limits_k {\left( {\frac{{\varepsilon _k^{ \sigma }  - \mu + M^{ \sigma } }}{{E_k^{ \sigma } }}\tanh \frac{{E_k^{ \sigma } }}{{2k_B T}}} \right)} } \right].
\label{eq_9}
\end{equation}
The Fock's parameter, $I_\sigma=\langle c_{i\sigma}^{+}c_{j\sigma}\rangle$, is given as

\begin{equation}
I_\sigma   = \frac{1}{2N}\sum\limits_k {\left (\frac{\varepsilon _k }{zt}\frac{\varepsilon _k^\sigma   - \mu + M^\sigma }{E_k^\sigma}\ {\rm tanh}\frac{{E_k^\sigma  }}{{2k_B T}}\right )}. 
\label{eq_10}
\end{equation}

\section{3. Numerical Results}
Solving self-consistently Eqs (\ref{eq_8})-(\ref{eq_10}) we obtain the phase diagrams showing the dependence of superconducting critical temperature, ferromagnetic critical temperature, and magnetic moment on pressure. In numerical calculations we have used $A = 0.013\;{\rm kbar}^{{\rm  - 1}}$ in Eq.\ (\ref{eq_3}), and the following relations for $S_1$ and $S_2$

\begin{equation}
S_1\equiv\frac{t_1}{t_0}=\frac{0.35}{1-0.003p},\;\;\;\;\;\;\;\;\;\; S_2\equiv\frac{t_2}{t_1}=\frac{0.25}{1-0.003p}. 
\label{eq_11}
\end{equation}

The relation (\ref{eq_11}) was assumed as the relation similar to the relation describing the pressure dependence of the hopping integral $t(p)$, Eq.~(\ref{eq_3}). The value $0.003$ in front of pressure was chosen as to obtain the Curie and superconducting critical temperatures in a good agreement with experimental data for ZrZn$_2$.

\begin{figure}[t]
\begin{center}
\epsfig{file=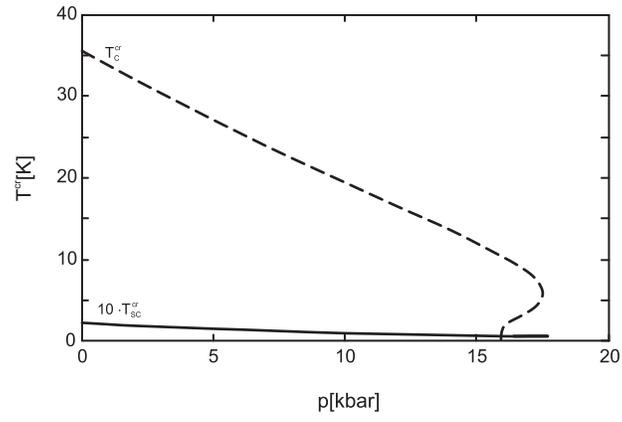,width=0.45\hsize}
	\end{center}
	\caption {ESP superconducting (solid line) and ferromagnetic (dashed line) critical temperature vs.\ pressure. The following values were used: $n=1.015$, $U=0.075\; {\rm eV}$, and $t_0=0.125\; {\rm eV}$.}
\label{fig_1}
\vskip0.5 cm
\end{figure}

Fig.~\ref{fig_1} shows the dependence of ESP superconducting, $T_{SC}^{cr}$, and ferromagnetic, $T_{C}^{cr}$, critical temperatures on external pressure. The value of electron occupation $n=1.015$ was chosen. This value allows obtaining the Curie temperature comparable with experimental data at $p=0\;{\rm kbar}$. The numerical results show that the Curie temperature at zero pressure is decreasing with growing carrier concentration (all remaining parameters being fixed). Note that the ESP superconducting critical temperature is magnified 10 times. As we can see, increasing the pressure causes decreasing of both superconducting and Curie temperatures. The pressure-dependence of critical temperatures is quasi-linear up to $p\approx16\;{\rm kbar}$. The effect of two values of Curie temperature at a given pressure above $16\;{\rm kbar}$ is caused by the specific shape of the density of states (logarithmic) and by the H-F approximation used in the model. The proper curvature of the Curie temperature dependence on pressure can be obtained by assuming semi-elliptic density of states or by use of the higher-order approximations for four- and six-operator terms in the Hamiltonian (e.g. Hubbard III) (see \cite{Mizia1}).

\begin{figure}[t]
\begin{center}
\epsfig{file=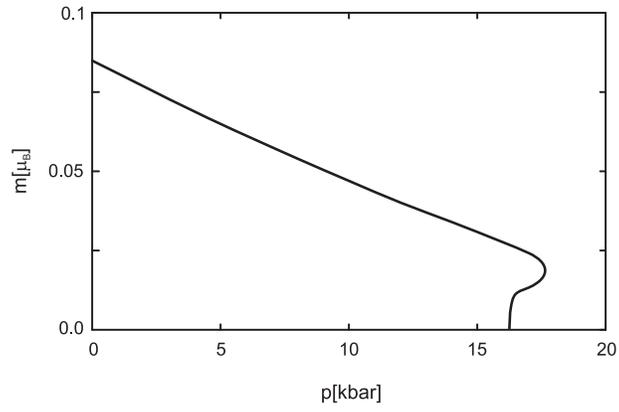,width=0.45\hsize}
	\end{center}
	\caption {Magnetic moment (in Bohr's magnetons) vs.\ pressure. The values used are the same as in Fig.~\ref{fig_1}, i.e.: $n=1.015$, $U=0.075\; {\rm eV}$, and $t_0=0.125\; {\rm eV}$.}
\label{fig_2}
\vskip0.5 cm
\end{figure}

Fig.~\ref{fig_2} shows the dependence of magnetic moment on external pressure. As we can see, the dependence $m(p)$ is also quasi-linear up to $p\approx16\;{\rm kbar}$, in good agreement with experimental data. The strange behavior of this curve above $16\;{\rm kbar}$ was explained earlier, in the description of Fig.~\ref{fig_1}.  

\section{Conclusions}
The simple itinerant electron model of Eq.\ (\ref{eq_1}) with additional assumption given by Eq.\ (\ref{eq_3}) and Eq.\ (\ref{eq_11}) can explain the pressure dependence of superconducting critical temperature, ferromagnetic critical temperature, and magnetic moment.

The abrupt decrease of magnetic moment at $p\approx16.5\; {\rm kbar}$ is related to the structural phase transition (see \cite{Uhlarz1}), which we do not take into account in our model.

It is worthwhile to note in here, that taking into account the inter-site correlations (given by the expression $\langle c_{i\sigma}^{+}c_{j\sigma}\rangle )$ gave us the nonzero values of parameters $S_1$ and $S_2$, and equivalently the nonzero values of parameters $\Delta t$ and $t_{ex}$. The last two are responsible for an existence of small superconductivity stimulated by ferromagnetism. Both superconductivity and ferromagnetism have pressure dependence in agreement with the experimental data.

\end{document}